\documentclass{article}

\usepackage{PRIMEarxiv}

\usepackage[utf8]{inputenc} % allow utf-8 input
\usepackage[T1]{fontenc}    % use 8-bit T1 fonts
\usepackage{hyperref}       % hyperlinks
\usepackage{url}            % simple URL typesetting
\usepackage{booktabs}       % professional-quality tables
\usepackage{amsfonts}       % blackboard math symbols
\usepackage{nicefrac}       % compact symbols for 1/2, etc.
\usepackage{microtype}      % microtypography
\usepackage{lipsum}
\usepackage{fancyhdr}       % header
\usepackage{graphicx}       % graphics
\graphicspath{{media/}}     % organize your images and other figures under media/ folder

\title{Abstracted Trajectory Visualization for Explainability in Reinforcement Learning
}

\author{
  Yoshiki Takagi, Roderick Tabalba, Nurit Kirshenbaum, Jason Leigh  \\
  Information \& Computer Science \\
  University of Hawaii at Manoa \\
  Honolulu\\
  \texttt{takagiyo@hawaii.edu} \\
}

\begin{document}
\maketitle

\begin{abstract}
Explainable AI (XAI) has demonstrated the potential to help reinforcement learning (RL) practitioners to understand how RL models work. However, XAI for users who do not have RL expertise (non-RL experts), has not been studied sufficiently. This results in a difficulty for the non-RL experts to participate in the fundamental discussion of how RL models should be designed for an incoming society where humans and AI coexist. Solving such a problem would enable RL experts to communicate with the non-RL experts in producing machine learning solutions that better fit our society. We argue that abstracted trajectories, that depicts transitions between the major states of the RL model, will be useful for non-RL experts to build a mental model of the agents. Our early results suggest that by leveraging a visualization of the abstracted trajectories, users without RL expertise are able to infer the behavior patterns of RL.
\end{abstract}

% keywords can be removed

\section{Introduction}
Explainable AI (XAI) research has shown the potential to close the gap between humans and RL models by providing explanations that help users understand how the black-box models work. So far, various aspects of explainability for users without RL expertise (non-RL experts) has been illuminated. These include explaining the policy of an agent~\cite{hayes2017improving, struckmeier2019autonomous}, justifying the action of an agent with reward~\cite{tabrez2019explanation, juozapaitis2019explainable}, and explaining the dynamics of an environment~\cite{elizalde2007mdp, khan2009minimal}. Recently, counterfactual explanations that explain "\emph{If A did not happen, B would not have happened}" and contrastive explanations that answer "\emph{Why $B_1$ rather than $B_2$?}" have been actively considered as good explanations that are understandable for a wide variety of users~\cite{van2018contrastive, olson2021counterfactual, madumal2020explainable}. 

However, the gap between RL practitioners and non-RL experts still needs much attention. As a result, this makes it difficult for non-RL experts to participate in the fundamental discussion of how RL models should be designed. For example, in conventional automobile development, a prototype vehicle developed by engineers is evaluated by test drivers with in-depth knowledge of driving, and modified by the engineers as necessary. However, discussions on the more fundamental question of how a self-driving car should be driven in the first place involves concerns about how machines will make moral decisions~\cite{awad2018moral}. Thus, such discussions should be opened to the general public. This indicates the need for a new XAI approach that requires less descriptive knowledge so that non-RL experts can build the mental model of RL models.

We argue that a visualization of abstracted trajectories, a visual representation of a behavior that consists of a trajectory that depicts transitions between the major states of the RL model, will be useful for non-RL experts to build a mental model of the agents. This is akin to estimating the personality of the driver by observing their driving performance from short video recordings. Thus, this approach is expected to provide a visual representation that efficiently summarizes agents’ behavior for non-RL experts. Based on this point of view, this short paper examines the question of \emph{how trajectories should be abstracted and visualized to support a user’s intuitive inference of the agent’s behavior}. Therefore, this study introduces a trajectory abstraction algorithm and proposes an interface for an abstracted trajectory visualization for various types of agents. Our preliminary observations from pilot study suggest that although the proposed interface has room for improvement, the abstracted trajectory visualization was useful for users without RL expertise to infer agents’ behavior. 

This short paper is structured as follows: the next section covers the technical background needed to understand an algorithm used to generate the abstracted trajectories; Section \ref{Section:Related Work} presents a literature review of relevant research; From section \ref{Section:aboutSystem} through \ref{Section:Discussion}, outline of proposed interface, preliminary evaluation, result, and discussions are provided; Finally, section \ref{Section:limitations} and \ref{Section:Conclusion} give limitations and a conclusion.

\section{Background}\label{section:Background}
\begin{figure}[ht]
\centering
\includegraphics[width=0.4\linewidth]{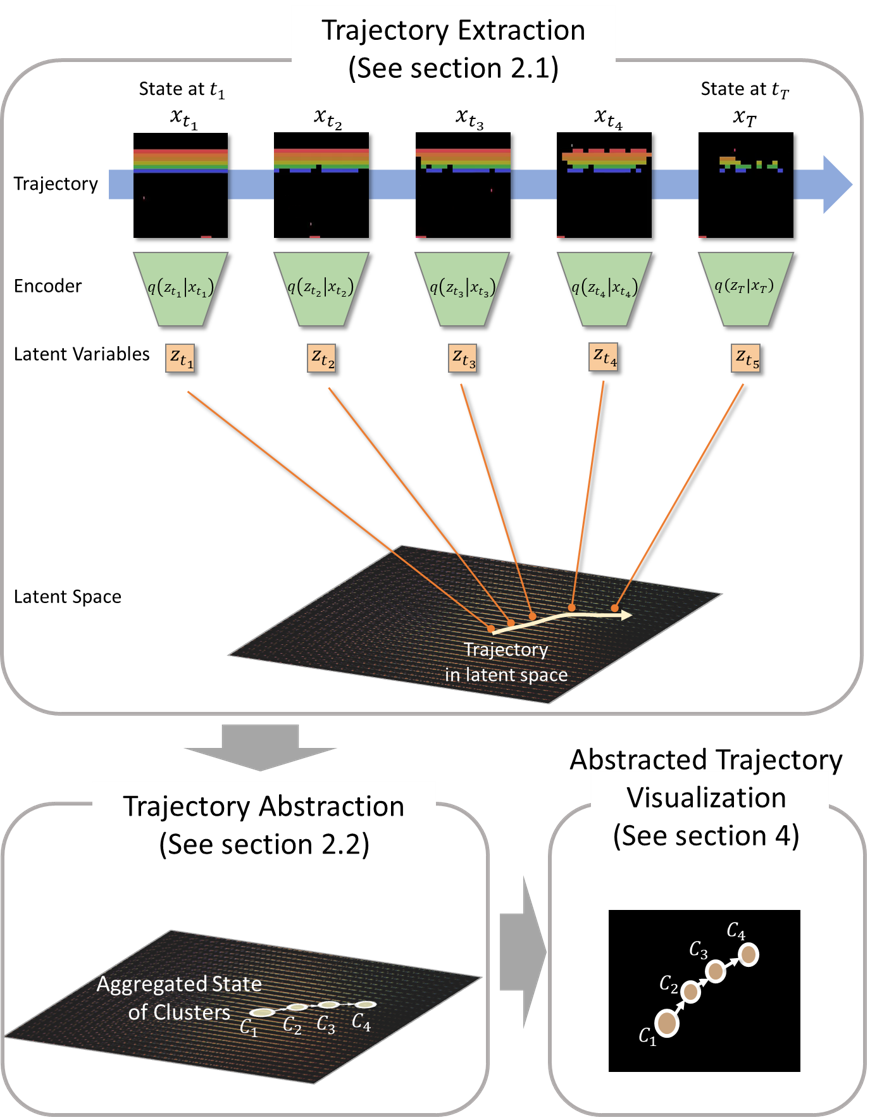}
\caption{The process of generating abstracted trajectories from input images that a RL agent observed. The features $z$ (latent variables) are extracted from the input images $x$ with VAE (See section \ref{subSection:Trajectory Extraction}). The extracted features $z$ are abstracted by spatio-temporal clustering (See section \ref{subSection:Trajectory Abstraction}).}
\label{fig:Figure1}
\end{figure}

This section provides technical background on machine learning models to extract transitions between the major states of the RL model. Please see section \ref{Section:aboutSystem} for how the extracted transitions are visualized. As shown in figure \ref{fig:Figure1}, the entire process of the algorithm has three steps: trajectory extraction, trajectory abstraction and abstracted trajectory visualization. In the following two subsections, the trajectory extraction and trajectory abstraction will be explained. 

\subsection{Trajectory Extraction}\label{subSection:Trajectory Extraction}
\begin{figure}[ht]
\centering
\includegraphics[width=0.7\linewidth]{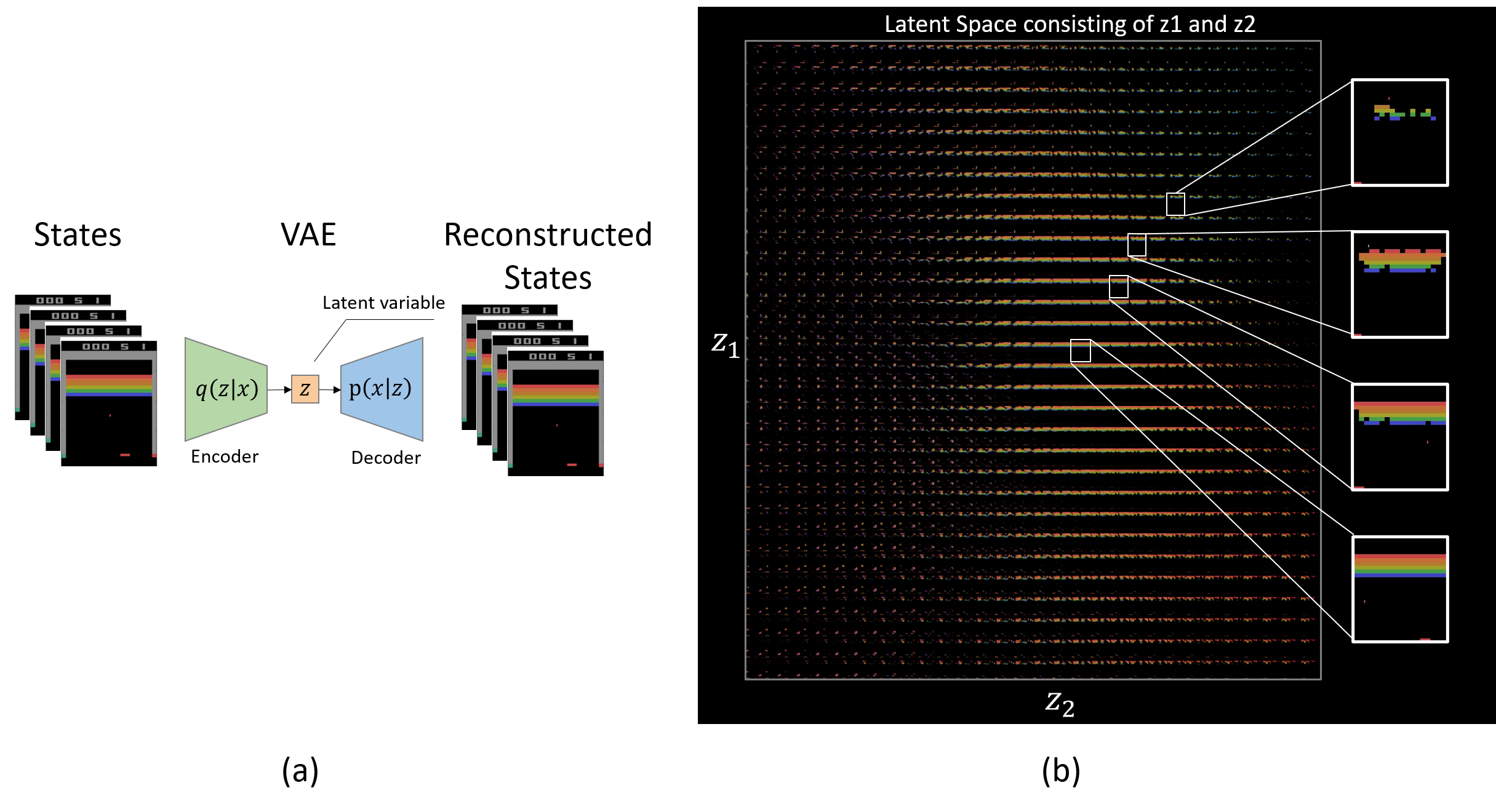}
\caption{a) The architecture of VAE, input images are fed to the decoder of VAE and the decoder is trained to compress inputs to latent variable $z$. The encoder of VAE is also trained at the same time to reconstruct the inputs from corresponding latent vectors. b) The reconstructed images projected on the latent space formed $z_1$ and $z_2$.}
\label{fig:Figure2}
\end{figure}

In this study, we used a machine learning model called Variational Autoencoder (VAE) to extract interpretable series of features without human supervision. Given an input item $x$, the goal of the VAE is to learn a vector $z$ that captures the features of $x$ in a disentangled manner (see Figure \ref{fig:Figure2} (a)). For example, when the VAE trained with images of Breakout, a game in which the player breaks rainbow-colored blocks by moving the paddle left and right and hitting the ball back to the block, the VAE learns to represent an input image of the game using $z$.  Since the dimension of the vector $z$ is usually high, dimensional reduction techniques, such as PCA~\cite{karl1901onlines}, t-SNE~\cite{vandermaaten08a} and UMAP~\cite{McInnes2018}, are used to project learned representation in 2D space. For instance, the feature of number of remaining blocks captured by $z$ can be visualized with PCA in a 2D space (see Figure \ref{fig:Figure2} (b)). Ideally these independent factors are expected to be captured in separated dimensions of $z$ (i.e., semantic dimensions), however, in real cases, some dimensions of $z$ will not learn to have semantic meanings. In order to have more semantic meaning in the latent vector, we deal with this problem by introducing $\beta$-VAE~\cite{higgins2016beta} which can facilitate $z$ to learn more disentanglement representation than vanilla VAE. The $\beta$-VAE is trained to minimize the following loss function:
\begin{equation}
\mathcal{L}_{\beta}(\theta, \phi | x)=-\mathbb{E}_{q_{\phi}(z|x)}[log \left(p_{\theta}(x|z) \right)] + \beta D_{\mathbb{KL}} \left(q_{\phi}(z|x)||p_{\theta}(z) \right)
\end{equation}
The first term is a reconstruction loss, and the second term is a regularization for disentanglement. With $\beta$ > 1, $\beta$-VAE encourages disentangled $z$ by putting a constraint on the latent bottleneck. Consequently, similar images are positioned closely together in the latent space, and as the state changes continuously due to the actions of the agent, the state transitions are represented as trajectories in the latent space. In this research, the architecture of the $\beta$-VAE follows the approach presented by Ha and Schmidhuber~\cite{ha2018recurrent}.

\subsection{Trajectory Abstraction}\label{subSection:Trajectory Abstraction}
In order to abstract the trajectories that the $\beta$-VAE generated, ST-DBSCAN~\cite{cakmak2021spatio} was used as a clustering algorithm, which is a density-based clustering technique designed for spatio-temporal data. Unlike many clustering methods that assume the independence and identically distributed (i.i.d.) assumption, ST-DBSCAN considers the temporal aspects. It forms a cluster when the number of points in a spatio-temporal region of specified radius around a point exceeds a specified threshold. This enables the clustered state transitions to capture the temporal structure inherent in the original data. 

ST-DBSCAN classifies the features $z$ that is obtained from the encoder of $\beta$-VAE into some clusters. Major states on a trajectory are generated by feeding the median of the distribution of $z$ for each cluster to the decoder of $\beta$-VAE. The abstracted trajectories are obtained by connecting these states in chronological order.

\section{Related Work}\label{Section:Related Work}
Our study is closely related to visual analytics (VA) for improving explainability of RL agents, and trajectory visualization for temporal data. This section describes the efforts in these two research areas and summarizes the differences between these efforts and our study.

The potential of finding hypotheses that can effectively explain the visualized behavior patterns of RL agents through interaction with an interface has been highlighted in the field of Visual Analytics (VA). These interfaces have been primarily designed for RL practitioners to discover behavior patterns and potential bugs of an agent~\cite{strobelt2017lstmvis, ming2017understanding, jaunet2020drlviz, wang2018dqnviz, mcgregor2015facilitating}. These methods enable the RL practitioners to construct accurate mental models about an agent by examining information in multiple charts about the agent. However, they remain inaccessible to users without such expertise. In addition, these methods cannot compare different types of machine learning models since their interfaces are designed for a single specific type of machine learning models such as RNN~\cite{medsker1999recurrent}, LSTM~\cite{hochreiter1997long}, and DQN~\cite{mnih2015human}.

Trajectory visualization can project the behavior patterns of agents as trajectories in single visual representation. For instance, Projection Path Explorer visualized several solutions of a Rubik’s Cube as trajectories and analyzed how these solutions were visualized as patterns of trajectories~\cite{hinterreiter2021projection}. The researchers who devised the Projection Path Explorer found that the characteristics of an agents’ behavior pattern can be expressed as clusters of states on the trajectories or bundles of the trajectories. Zahavy \emph{et al}. have proposed an interface that visualizes latent vectors of a RL model in two dimensions using t-SNE~\cite{zahavy2016graying}. They provided case studies to demonstrate how users can read agent behavior patterns from the trajectories of the latent vectors in the interface. Both studies have provided case studies to demonstrate how an agent's behavior patterns can be read from trajectory visualizations, however, they did not conduct user studies to evaluate how users actually use the interface to find agent’s behavior patterns. Furthermore, these interfaces are difficult for non-RL experts to use, because their methods directory project high-dimensional states into 2D space using dimensionality reduction techniques such as t-SNE, resulting in the visualized trajectories are often complex.

In this short paper, we propose a new trajectory abstraction algorithm to reduce the complexity of trajectories of different types of RL models so that users without RL expertise can easily see a high-level view of RL agents’ behavior patterns. We also conduct a pilot study to evaluate the user’s comprehension when observing the abstracted trajectory visualization.

\section{TRAJECTORY VISUALIZATION INTERFACE}\label{Section:aboutSystem}
\begin{figure}[ht]
\centering
\includegraphics[width=0.8\linewidth]{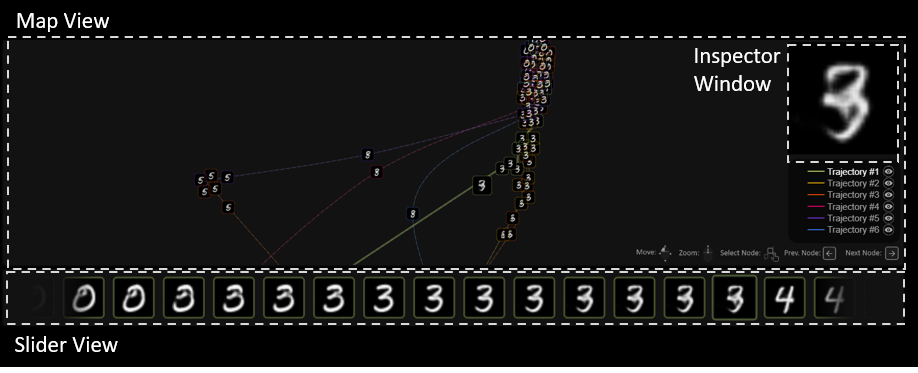}
\caption{Proposed interface for abstracted trajectory visualization: A map view is placed in the center of the interface, and a slider view is placed on the bottom. Enlarged image of a hovered over node is displayed in the inspector window on the top right.}
\label{fig:Figure3}
\end{figure}

This section describes an interactive interface that visualizes abstracted trajectories. The interface (see Figure \ref{fig:Figure3}) consists of two parts: a map view drawing the trajectories as a directed graph, and a slider view visualizing the trajectories horizontally. In the following sub-sessions, the details of the map view and slider view will be described.

\subsection{Map View}
The map view is located in the center of the interface and visualizes the abstracted trajectories as a directed graph. Nodes of the directed graph represent major states that were obtained from the clusters using ST-DBSCAN as explained in Section \ref{section:Background}. Position of the nodes in the map is determined by a force simulation of D3JS~\cite{bostock2011d3}. By applying the Euclidean distances of the latent vectors $z$ corresponding to the major states to the strength of link between nodes, the force simulation can form clusters based on the distance of the major states. Edges of the directed graph are visualized based on the temporal dependencies between the nodes and are drawn with Bézier interpolation to improve readability. The flow of time is visualized as an animation of dots drawn on the edges. 

Users can interact with the trajectories on the map by scrolling, dragging, hovering, and clicking with a mouse. Scrolling and dragging will move and zoom the trajectories on the map like a Google Map. During exploration, users can hover over nodes in the map to see the enlarged image of the node displayed in the inspector window in the upper right corner. By tracing nodes along a trajectory, users can also examine state transitions along the trajectory displayed in the inspector window. Finally, by clicking on a node, the user can highlight the trajectory to which the clicked node belongs. By hovering over a node belonging to another trajectory, the user can interactively compare the shape of the highlighted trajectory that is clicked with the trajectory that is hovered. The highlighted trajectory that is clicked is also reflected in the slider, which will be described in detail in the next section.

\subsection{Slider View}
The slider view projects nodes of abstracted trajectories horizontally in a chronological order. Since the nodes in the slider view are sorted from left to right, users can easily see state transitions on the abstracted trajectories by comparing the adjacent images.

Like the map view, the users can interact with the nodes in the slider view by scrolling, hovering, and clicking with a mouse. The users are able to explore the slider view by using the scroll wheel, which will scroll the slider view horizontally. Also, by hovering over a node in the slider, the users can see the enlarged image of the node in the inspector window on the top right.  Finally, by clicking on a node, the trajectory belonging to that node is highlighted on the map. This allows the users to see the shape of the trajectory in the map view and the state transitions in the slider view at the same time.

\section{PRELIMINARY EVALUATION}\label{Section:Evaluation}
\begin{figure}[ht]
\centering
\includegraphics[width=0.8\linewidth]{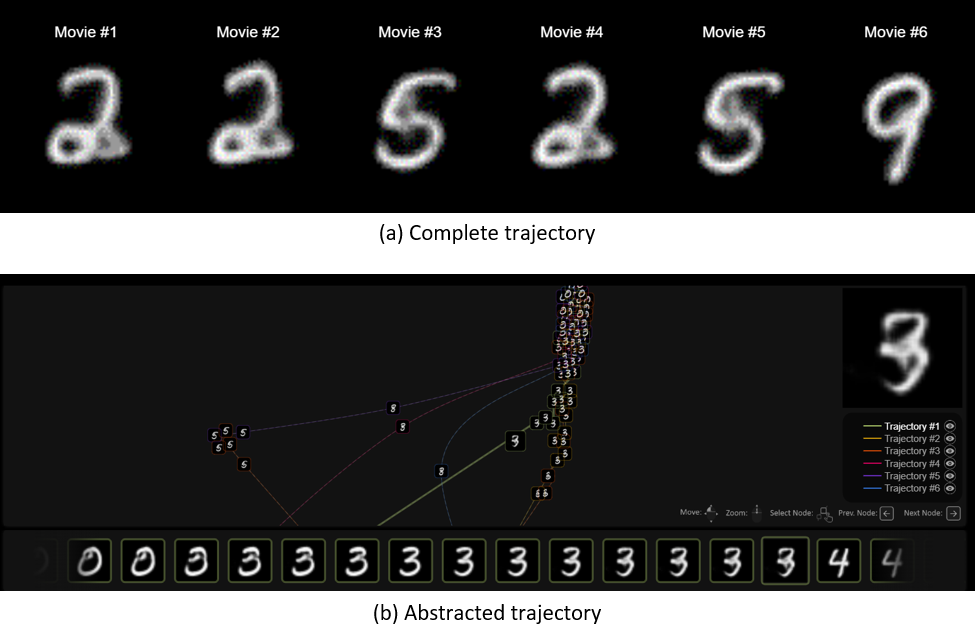}
\caption{Two visualization types used in a comparative evaluation}
\label{fig:Figure4}
\end{figure}

We performed a preliminary comparative evaluation between two visualization types, with/without trajectory abstraction. In the case of without the abstraction, animations as shown in figure \ref{fig:Figure4} (a) are presented to participants. We call this case “complete trajectory” in this short paper since the animations can be understood as the visualization of complete state transitions. In the second case, which includes the trajectory abstraction as explained in the previous section (See Figure \ref{fig:Figure4} (b)) is provided to participants. In both cases, we measured an accuracy to complete a task of which details will be explained in section \ref{subsection:task}. This pilot study was designed to be conducted online so that a wide range of participants can participate in this study in the future. 

\subsection{Questions}\label{subsection:questions}
In order to evaluate whether the abstracted trajectories are useful for non-RL experts to build a mental model of the agents' behavior, the following three questions were set to guide this pilot study:
\begin{itemize}
\item[Q1-] How well does a user’s mental model obtained from the abstracted trajectory agree with a mental model obtained from complete trajectory?
\item[Q2-] Is the trajectory abstraction algorithm able to extract information that helps users to understand the behavioral pattern of RL agent?
\item[Q3-] What kind of information do users read from the abstracted trajectories?
\end{itemize}

\subsection{Task}\label{subsection:task}
\begin{figure}[ht]
\centering
\includegraphics[width=0.8\linewidth]{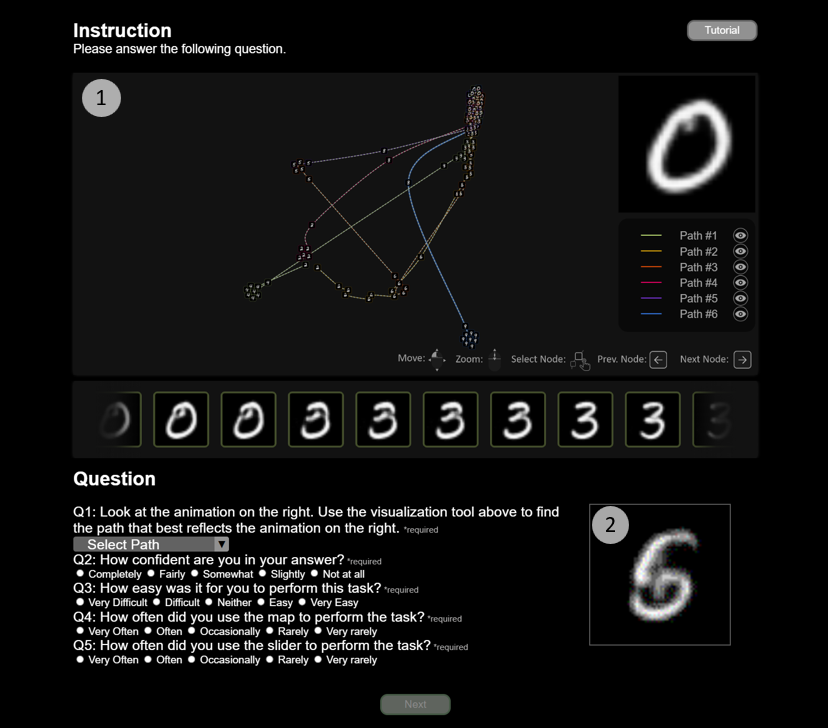}
\caption{Designed analytical task in the case of abstracted trajectory: Participants are asked to find a trajectory visualized in the interface of abstracted trajectory (see region (1) in the figure) that matches an animation (see region (2) in the figure). In the case of complete trajectory, the interface in the region (1) is replaced with visualizations of complete trajectory shown in figure 4 (a). }
\label{fig:Figure5}
\end{figure}

To evaluate the performance of each visualization type, we created a short analytical task as shown in figure \ref{fig:Figure5}. The task requires participants to observe an animation changing over time and to identify the trajectory that matches the animation from six different trajectories represented by either complete trajectories or abstracted trajectories. The aim of the task is to measure how correctly the participants are able to generalize an agent’s policy from each visualization type.

We designed the task for 6 different applications in total as shown in table \ref{tab:table1}. Of the 6 applications, an application of Mnist is used for a tutorial which is presented to the participants at the beginning of the pilot study. The tutorial displays animations of handwritten digits of 0-9. The animation includes a smooth transition of the handwritten digits. As an example, for the pattern 0$\rightarrow$4$\rightarrow$3$\rightarrow$2, the animation would initially show 0, then slowly morph into a 4, and so on. In the case of the complete trajectory, the participant needs to perform the task by looking at six different animations (see Figure \ref{fig:Figure4} (a)).  In the case of abstracted trajectory, they need to use an interface shown in figure \ref{fig:Figure4} (b) to identify which trajectories match the animation.

The five applications except for Mnist were used after the tutorial. Thus, a total of 10 questions (2 visualization types $\times$ 5 applications) were randomly given to the participants for an actual test. The details of the procedure after the tutorial will be explained in section \ref{subsection:procedure}. 

Dataset of state transitions used in each application were obtained from the work by Such \emph{et al}.~\cite{such2018atari}. The research group trained six RL models including A2C~\cite{mnih2016asynchronous}, ApeX~\cite{horgan2018distributed}, DQN~\cite{mnih2015human}, ES~\cite{salimans2017evolution}, GA~\cite{such2017deep}, and Rainbow~\cite{hessel2018rainbow}, on various Atari games to support research that investigates the properties of these agents. In our study, replays of an Atari game played by these six RL models were represented as six trajectories. A participant never encountered a task with the same answer across visualization types. 

In summary, we used 2 visualization types (with/without trajectory abstraction) within subject design. The blocks of questions are counterbalanced to avoid bias and participants never encountered a task with the same answer across visualization types. Thus, the total of trials including the tutorial result in 12$N$ trials (2 visualization types $\times$   6 applications including tutorial $\times$ $N$ participants). 

\begin{table}[ht]
  \caption{Applications used in this pilot study}
  \label{tab:table1}
  \begin{tabular}{cccccc}
    \toprule
    Mnist&Breakout&Qbert&Amidar&SpaceINvader&Boxing\\
    \midrule
    \raisebox{-\totalheight}{\includegraphics[width=0.1\linewidth, height=0.1\linewidth]{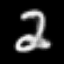}}
    & 
    \raisebox{-\totalheight}{\includegraphics[width=0.1\linewidth, height=0.1\linewidth]{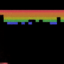}}
    &
    \raisebox{-\totalheight}{\includegraphics[width=0.1\linewidth, height=0.1\linewidth]{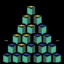}}
    &
    \raisebox{-\totalheight}{\includegraphics[width=0.1\linewidth, height=0.1\linewidth]{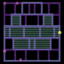}}
    &
    \raisebox{-\totalheight}{\includegraphics[width=0.1\linewidth, height=0.1\linewidth]{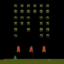}}
    &
    \raisebox{-\totalheight}{\includegraphics[width=0.1\linewidth, height=0.1\linewidth]{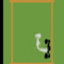}}\\
  \bottomrule
\end{tabular}
\end{table}

\subsection{Procedure}\label{subsection:procedure}
In the pilot study, a total of 12 questions (2 visualization types $\times$ 6 applications including tutorial) were asked for each participant. The participants were asked to complete each question with unlimited time and they could not redo questions. We measured accuracy of response for performance evaluation. Accuracy of each visualization type was measured by the number of correct responses out of the total number of responses in each application. The comparison of participants’ accuracy between the two visualization types can be understood as a measure of how correctly the participants are able to generalize an agent’s policy from abstracted trajectories compared to complete trajectories. At every question, participants were asked to rate their confidence on their answer and the difficulty of the question subjectively with a 5-point Likert scale. For the questions that included the abstracted trajectory, we additionally asked them to rate how frequently they used map view and slider view in answering questions with a 5-point Likert scale to examine how they used the interface of the abstracted trajectory.

After completing all questions, participants were asked to rate subjectively about the ease of use and usefulness of the abstracted trajectory interface, using a 5-point Likert scale. We also provided some open-ended questions focusing on what they could read from the interface of abstracted trajectory and what they found difficult to read from the interface of abstracted trajectory. 

\subsection{Experimental Setup}
All experiment sessions were conducted online. Participants used their own laptop or desktop to complete the task. The window size was recorded during the pilot study to exclude participants who conducted the task with too small window size. As a result of screening, no participants were excluded. The pilot study was approved by our internal IRB. On average, the study lasted about 45 min. 

\subsection{Participants}
\begin{figure}[ht]
\centering
\includegraphics[width=0.6\linewidth]{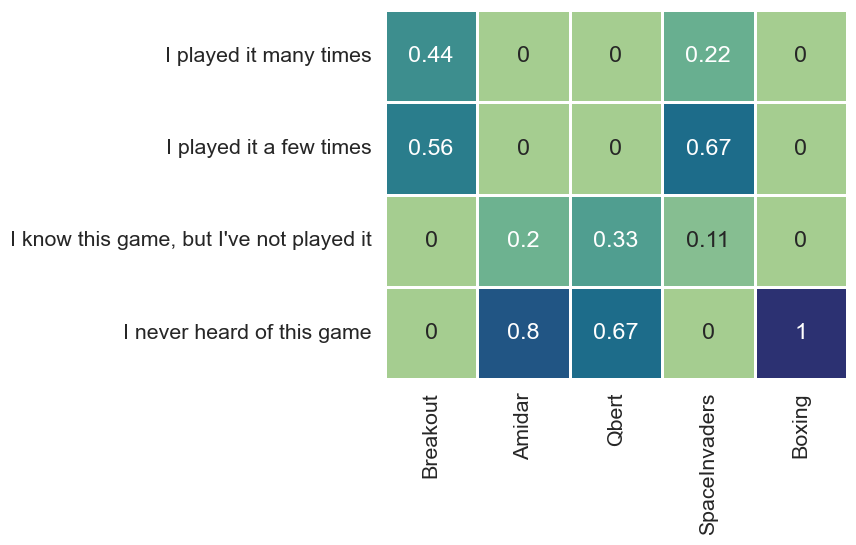}
\caption{Participants’ familiarity of applications used in the pilot study. The horizontal axis is the types of applications. The vertical axis is the rating of 4-point Likert scale}
\label{fig:Figure6}
\end{figure}

We recruited 9 participants (5 male and 4 female) who are non-RL experts, aged between 22 and 45 years, from computer science. The participants rated the familiarity of the applications used in the study using a 4-point Likert scale.  As shown in figure \ref{fig:Figure6}, most participants answered they have played the game at least once in Breakout and Space Invaders. Other than the two games, all participants self-declared as not so familiar about the games. 

\section{Results}\label{Section:Result}
This section presents quantitative (accuracy) and qualitative (subjective ratings and responses of open-ended questions) results. We used Fisher's exact test ($\alpha = .05$) for accuracy comparisons and Mann-Whitney U test ($\alpha = .05$) for other comparisons.

\subsection{Quantitative Results}
\begin{figure}[ht]
\centering
\includegraphics[width=0.5\linewidth]{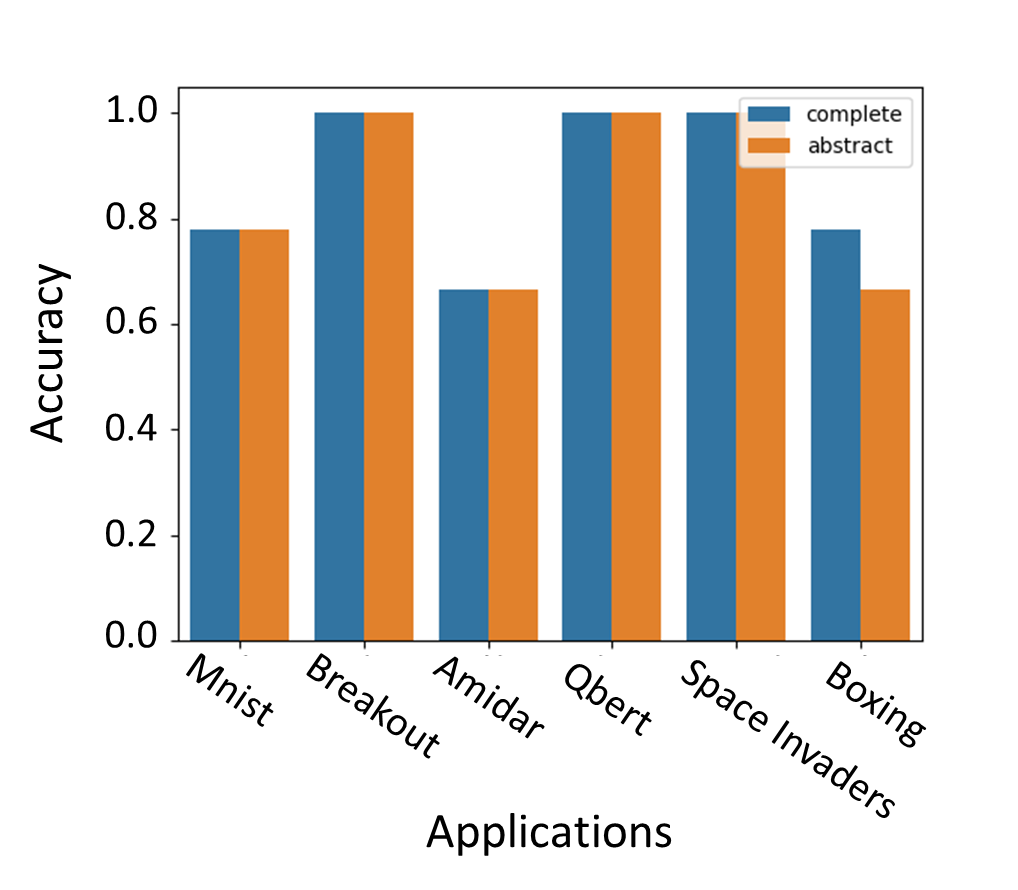}
\caption{Results of Accuracy: No significant differences were found between the two groups in all applications.}
\label{fig:Figure7}
\end{figure}

Accuracies between abstracted and complete trajectories were shown in figure \ref{fig:Figure7}. We used Fisher's exact test for accuracy comparisons between the two groups, since some cell frequencies are less than 5. The test results show no statistical differences were found between the two groups.

\subsection{Qualitative Results}
\subsubsection{Confidence and Difficulty of the task}

\begin{figure}[ht]
\centering
\includegraphics[width=0.5\linewidth]{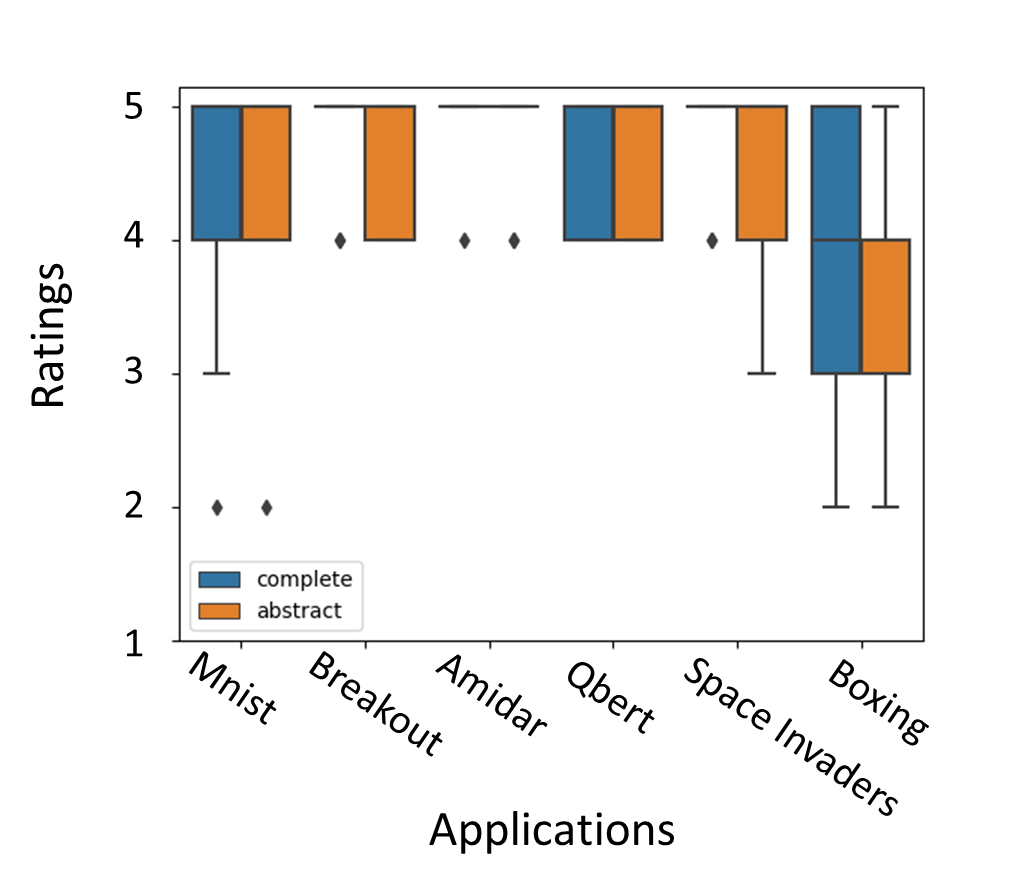}
\caption{Results of Confidence: The vertical axis is the rating of 5-point Likert scale about participants’ confidence on their answer (1 - “Not at all”, 2 - “Slightly”, 3 - “Somewhat”, 4 - “Fairly” and 5 - “Completely”).}
\label{fig:Figure8}
\end{figure}

\begin{figure}[ht]
\centering
\includegraphics[width=0.5\linewidth]{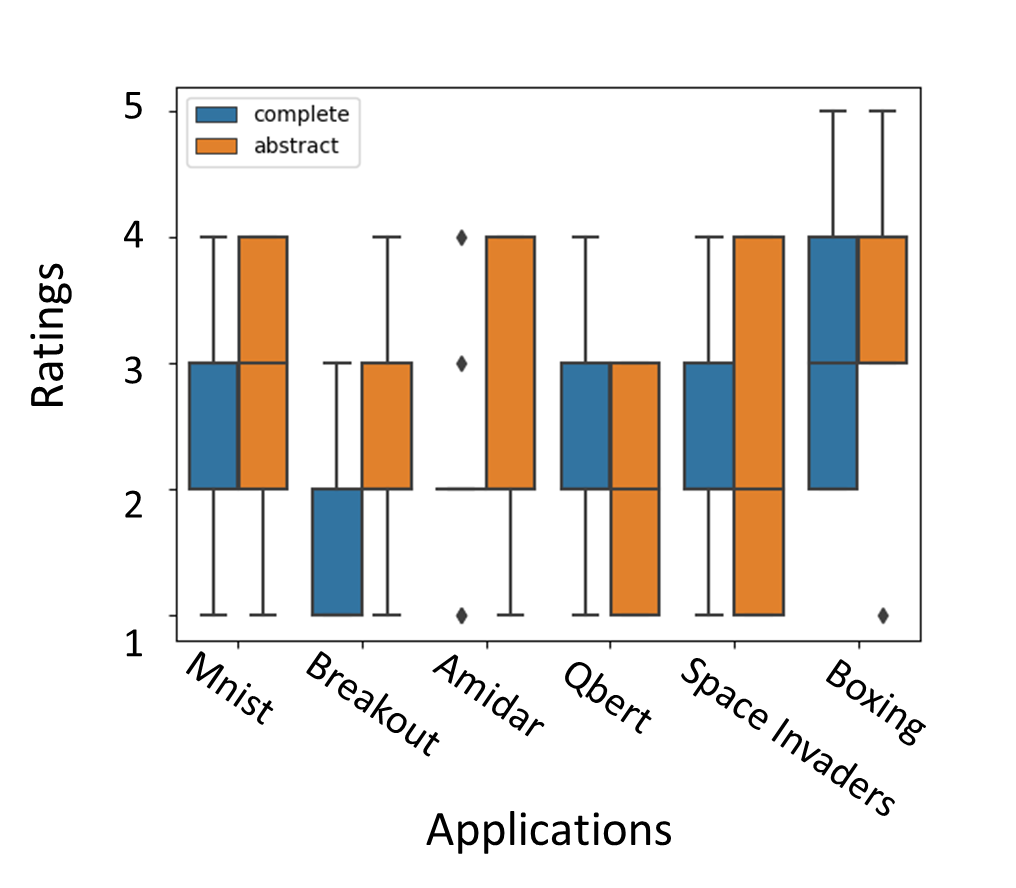}
\caption{Results of Difficulty: The vertical axis is the difficulty rating of 5-point Likert scale about the task (1 - “Very easy”, 2 - “Easy”, 3 - “Neither”, 4 - “Difficult” and 5 - “Very difficult”).}
\label{fig:Figure9}
\end{figure}

Participants’ confidence on their answer and difficulty ratings of the task between abstracted and complete trajectories are shown in figure \ref{fig:Figure8} and \ref{fig:Figure9} respectively.  As a result of Mann-Whitney U test, no significant differences of the confidence and the difficulty rating were found between the two groups.

\subsubsection{Usefulness of the interface}
In this section, we will report the comments and feedback that we got from the participants after they have completed the study. As a reminder, the participants were asked to answer a question “Do you think the trajectory visualization tool is more useful than just watching animations for performing tasks?” with a 5-point Likert scale and write the reason for the answer. 

\begin{figure}[ht]
\centering
\includegraphics[width=0.7\linewidth]{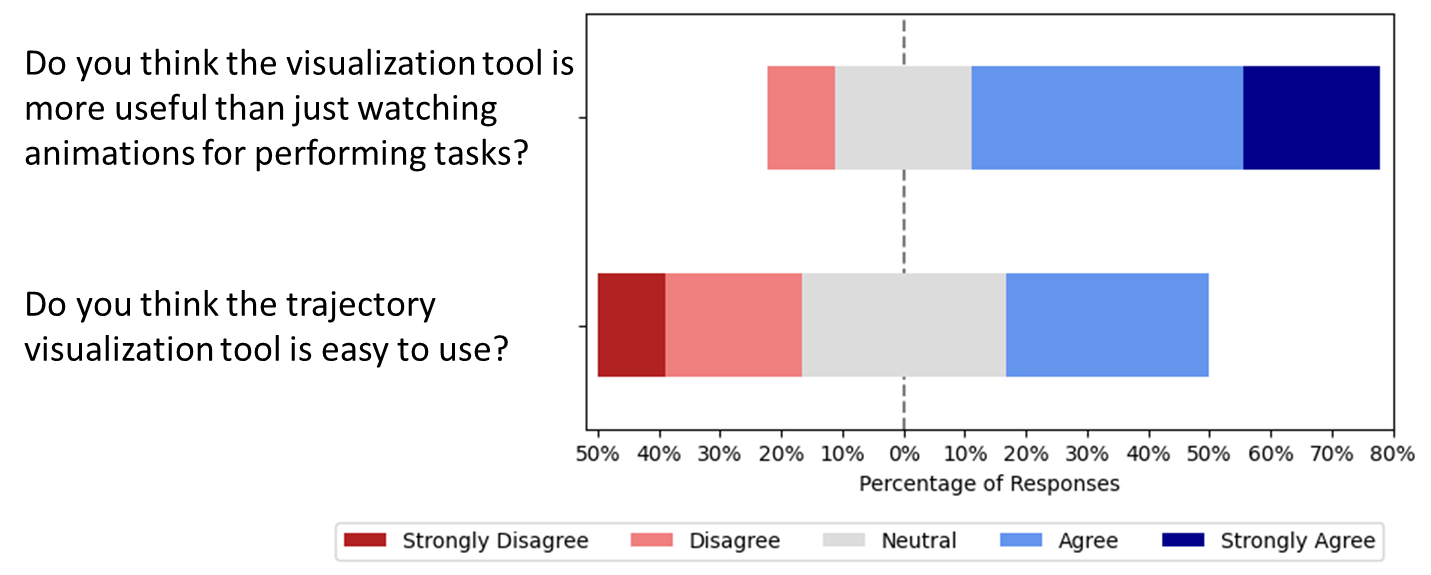}
\caption{Results of Usefulness and Ease of Use.}
\label{fig:Figure10}
\end{figure}

The majority of participants agreed that the interface for abstracted trajectory was useful (see Figure \ref{fig:Figure10}). However, opinions on the map view of the interface were divided as such: about half the participants reported the interface as generally useful, whereas the other felt not. A participant who answered neutral in the usefulness rating with the Likert scale said “(I am) not sure the map is for someone that has never dealt with data visualization.” A participant who answered disagree suggested that “it would be nice to visualize nodes in the map with less cluttered, because tight clusters do not help clarity.” Another participant who agreed on the rating scale left a positive comment about the map that “It helps compare and contrast what is similar and what is not.”

\begin{figure}[ht]
\centering
\includegraphics[width=0.8\linewidth]{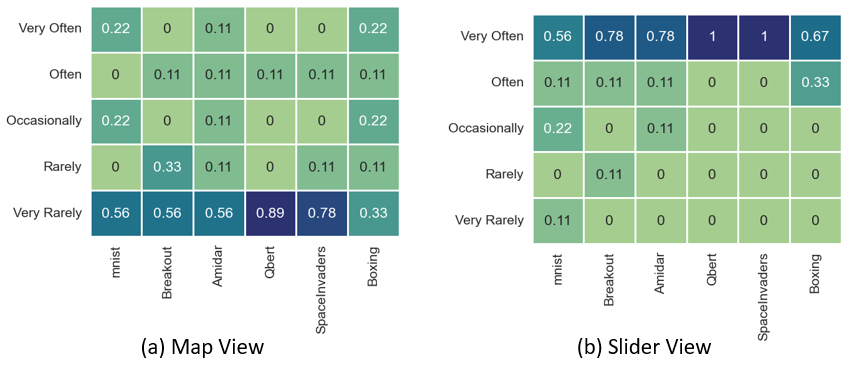}
\caption{Results of Usage: The vertical axis is the rating of 5-point Likert scale about frequencies of map and slider use. Participants use slider more frequently than map for performing tasks.}
\label{fig:Figure11}
\end{figure}

From the comment on the slider view, it seems that most participants thought that the slider was useful for performing tasks (see Figure \ref{fig:Figure10}). For an open-ended question “What did you find most useful about the tool over just watching the animation?”, the majority of participants mentioned the usefulness of the slider. For example, a participant stated that “The slider was really useful because it meant you could focus on certain parts of the animation.” This trend in comments is consistent with the tendency of participants to use the slider more frequently than the map to perform the task, as shown in Figure \ref{fig:Figure11}.

\subsubsection{Ease of use}
In this section, we will report the comments on the ease of use on the interface that we got from the participants after they have completed the study. As a reminder, the participants were asked to answer a question “Do you think the trajectory visualization tool is easy to use?” with a 5-point Likert scale. We also showcase their answer to a question “What was most difficult to understand from the visualization tool?”.

Participants were divided on whether the interface was easy to use, with an equal number of participants saying it was more or less neutral. Those who said it was easy to use explained that " It is not easy to use for finding a matching path but easy to use for eliminating similar paths.” A participant who answered neutral stated that “(It) was hard to follow the map and at the same time compare it with the animations”. A participant who strongly disagreed stated, "I did not find any purpose of using the map. Instead, I just toggled the path number on the right with the eye icons, then skimmed through the nodes to compare to the original animation." Overall, participants mentioned the ease of use of the map as a criterion for evaluating the ease of use of the interface.

\section{Discussions}\label{Section:Discussion}
The main goal of our evaluation was to evaluate whether visualizing abstracted trajectories can aid non-RL experts to build a mental model of the agents’ behavior. As a result of the pilot study, the accuracy scores of the trajectory identification task between abstracted and complete trajectory was comparable, and the interface of abstracted trajectory was subjectively preferred by the participants. These preliminary results seem to affirm $Q1$ and $Q2$ presented in section \ref{subsection:questions}. However, with respect to the map view in the interface, some users questioned its usefulness and ease of use. This suggests the need for improvements of the map view of the interface. In the following subsections, we will discuss the results of the pilot study to answer the questions $Q1$, $Q2$ and $Q3$ presented in section \ref{subsection:questions}, and point out some suggestions and improvements in designing an interface for the abstracted trajectories. 

\subsection{How well does a user’s mental model obtained from the abstracted trajectory agree with a mental model obtained from complete trajectory?}\label{subsection:discussionQ1}
The accuracies of the task using the abstracted trajectories were comparable to that of complete trajectories. It suggests that the trajectory abstraction algorithm can effectively distill significant information about the agent's behavioral patterns. A participant supported the effectiveness of the trajectory abstraction by saying that "You can see snapshots, you don't need to wait for the animation to loop." Another participant stated the benefit of abstracted trajectory on the visual memory that "Being able to go node by node, frame by frame, to compare the original animation to the different paths, is a lot easier than having to watch 6 other animations and comparing it to the original animation." This comment emphasizes the advantage in visual memory that abstracted trajectory visualization allows for at glance comprehension of state transitions, whereas looking at complete trajectory necessitates refreshing their visual memory. 

\subsection{Is the trajectory abstraction algorithm able to extract information that helps users to understand the behavioral pattern of RL agent?}
While discussions on section \ref{subsection:discussionQ1} suggest that the trajectory abstraction algorithm works to a degree, the comment of a participant that "It took a bit to get used to because many frames were omitted. You have to make a mental leap to see how the clusters could represent the state of each game." indicates a mismatch of trajectory abstraction between humans and the abstraction algorithm. The discrepancy between the spatial information learned by machine learning models and the concept learned by humans has been long discussed in the field of knowledge abstraction and representation learning~\cite{tenenbaum2011grow, lynn2020abstract}. And the discussion of how to design an interface to close such gaps is also still relatively new in the CHI domain~\cite{wang2023drava}. 

While revealing the mechanism of how we abstract a trajectory of a RL agent is esoteric, supervised learning and transfer learning may be useful for the trajectory abstraction algorithm to generate abstracted trajectories that are more similar to human intuition. For example, in the field of video summarization, where ML models generate short versions of original full-length video, various models using supervised learning have been proposed~\cite{bora2018review, haq2020video}. Thus, it would be interesting to apply transfer models used in this field to the trajectory abstraction in RL.

\subsection{What kind of information do users read from the abstracted trajectories?}
Arrangement of clusters of the nodes and trajectory shapes in the map become an important clue for users to identify a trajectory that is closer to an animation. For example, a user who analyzed trajectories commented, "It helps compare and contrast what is similar and what is not. You can also refer back and forth to see which one looks more similar and keep it in mind or uncheck it so that you eliminate it as an option.” Another user referred to the benefit of the clusters of the nodes, "clusters that were nicely isolated from others were easier to focus on finding the "why are they far away"". Because they read such information, the map would be a more useful tool when performing tasks such as grouping similar behavior patterns of RL agents or counting the number of agents visiting similar states. 

\subsection{How should we improve the interface?}
Comments from the pilot study revealed that the map view of the interface requires a refined interaction and a view organization. Regarding interaction, some users found it challenging to trace a trajectory in the map with a mouse cursor. New interaction technique designed for trajectory visualization such as direct manipulation is a promising solution~\cite{kondo2014dimpvis}. Such a technique is expected to allow users to navigate nodes along a trajectory more intuitively. Another issue is the overlap of nodes and trajectory, leading to hinder visibility and discourage users to use the map. Hence, applying clustering techniques to the position of the nodes based on Euclidean distances such as the k-means clustering is considered to be beneficial to make trajectory visualization less cluttered. 

Lastly, emphasizing temporal structures of the trajectory visualization in the map view may lead to user’s better understanding for the visualization. Majority of participants in the pilot study preferred the slider rather than the map. This may be because the slider, of which nodes align from left to right based on temporal dependency, facilitates an intuitive grasp of state transitions. Thus, a visualization incorporating a branching tree structure could be advantageous for performing the task. 

\section{Limitations and Future Work}\label{Section:limitations}
Because of the small sample size of the pilot study, many of the results presented in this short paper need to be treated as preliminary results. In addition, since all participants in the pilot study have a background of computer science, it will be necessary to receive evaluations from people with other backgrounds in the future. However, since the proposed evaluation framework is designed to be conducted online, we believe that crowdsourcing can be used to address these issues in the future.

\section{Conclusion}\label{Section:Conclusion}
In order to explain the behavior patterns of RL agents to non-RL experts in a non-descriptive way, this short paper introduced a XAI algorithm for generating abstracted trajectories from various RL agents’ trajectories and proposed the interface for the abstracted trajectory visualization. We designed an online evaluation framework for the interface to evaluate whether the abstracted trajectories are useful for non-RL experts to build a mental model of the agents. Pilot studies based on the framework suggest that the interface was useful for non-RL experts in the task of identifying agent behavior patterns. The evaluation framework also revealed that users preferred the slider to the map, and provided valuable insights and feedback for improving the visualization of the abstracted trajectories.

%Bibliography
\bibliographystyle{unsrt}  
\bibliography{references}

\end{document}